\documentclass{sig-alternate-05-2015}      
\usepackage{graphicx} 
\usepackage{color} 
\usepackage{amsmath} 
\usepackage{amssymb,amsfonts,amsmath}
\usepackage{algorithmic} 
\usepackage{url}
\usepackage{letltxmacro}
\usepackage{tikz}
\usepackage{caption}
\usepackage{subcaption}
\usepackage{enumitem}
\usepackage{listings}

\DeclareGraphicsExtensions{.pdf,.png}

\makeatletter
\LetLtxMacro\latexincludegraphics\includegraphics
\@ifpackageloaded{tex4ht}
	{\renewcommand{\includegraphics}[2][]{%
		\immediate\write18{s=\Ginput@path; (IFS='{}'; for i in $s; do convert ${i}#2 ${i}#2.eps; done)}%
		\latexincludegraphics[#1]{#2.eps}}}
	{\renewcommand{\includegraphics}[2][]{\latexincludegraphics[#1]{#2}}}
\makeatother

\renewcommand{\thefootnote}{*}

 \long\def\symbolfootnote[#1]#2{\begingroup%
\def\thefootnote{\fnsymbol{footnote}}\footnote[#1]{#2}\endgroup}

\usepackage{amsthm}

\newlength\Origarrayrulewidth

\newif\ifGPblacktext 
\GPblacktexttrue

\global\copyrightetc{This paper is supplementary to: \url{https://arxiv.org/abs/1611.04167}}

\begin{document}

\conferenceinfo{This paper is supplementary to:}{https://arxiv.org/abs/1611.04167}

\title{A parallel workload has extreme variability in a production environment}

\numberofauthors{4}

\author{%
\alignauthor Richard Henwood\\
\affaddr{Arm, Inc, 5707 Southwest Parkway, Austin, Texas, USA}\\
\email{richard.henwood@arm.com}
\alignauthor
N. W. Watkins\\
\affaddr{Centre for the Analysis of Time Series, London School of Economics, Houghton Street, London, UK}\\
\affaddr{Centre for Fusion, Space and Astrophysics, University of Warwick, UK}\\
\affaddr{Faculty of Science, Technology, Engineering \& Mathematics, Open University, Milton Keynes, UK}
\alignauthor
S. C. Chapman\\
\affaddr{Centre for Fusion, Space and Astrophysics, University of Warwick, UK}
\and
\alignauthor
R. McLay
\affaddr{Texas Advanced Computer Center, 10100 Burnet Rd, Austin, Texas, USA}
}

\maketitle

\begin{abstract}

Writing data in parallel is a common operation in some computing
environments and a good proxy for a number of other parallel processing
patterns. The duration of time taken to write data in large-scale compute
environments can vary considerably. This variation comes from a number of
sources, both systematic and transient. The result is a highly complex
behavior that is difficult to characterize. This paper further develops
the model for parallel task variability proposed in the paper ``A
parallel workload has extreme variability'' (Henwood et. al 2016). This
model is the Generalized Extreme Value (GEV) distribution.  This paper
further develops the systematic analysis that leads to the GEV model with
the addition of a traffic congestion term.  Observations of a parallel
workload are presented from a High Performance Computing environment
under typical production conditions, which include traffic congestion.
An analysis of the workload is performed and shows the variability tends
towards GEV as the order of parallelism is increased. The results are
presented in the context of Amdahl's law and the predictive properties of
a GEV models are discussed. A optimization for certain machine designs is
also suggested.

\end{abstract}

\section{Introduction}

Performance models of high performance computer (HPC) systems enable
architects and administrators to gain insight into optimal system
configuration for their applications. Models also provide valuable
comparison against measured values that can be used to evaluate the
behavior of system components (for example network, storage nodes). A
common approach is to directly observe a mean value for the property of
interest (bandwidth or latency for example) and with domain expertise,
historical knowledge, and instantaneous system insight, judge if the
measured value is tolerable.

In the paper ``A parallel workload has extreme variability''
\cite{henwood16} it is claimed that the duration of time taken for a
sufficiently parallel workload to complete should follow the Generalized
Extreme Value (GEV) distribution shown in Eqn \ref{eqn:gevpdf}. A parallel
write makes a good choice for study as it is common, well understood, and a
good proxy for other common parallel operations, including map-reduce, and
sharded databases.

\begin{equation}
P_{\mu,\sigma,\xi}(x) =
\begin{cases}
\exp \left( - \left( 1+\xi \left(\frac{x-\mu}{\sigma}\right)\right)^{-1/\xi} \right) & \text{if} 1+\xi\frac{(x-\mu)}{\sigma} > 0 \\

\exp \left( - e^(\frac{x-\mu}{\sigma})\right) & \text{if} \xi = 0
\end{cases}
\label{eqn:gevpdf}
\end{equation}

Extreme value theory and significant observations of the GEV across many
scientific disciplines. The modern theory can be traced back to Fisher and
Tippett in the 1920s \cite{fishertippett1928}. A brief description of their
discovery and recent important observations of extreme phenomena are recorded
in \cite{henwood16} and references therein.

The argument presented for GEV in parallel environments is, in essence,
that the parallel task is executed by workers that are completely
independent with identical distribution (IID). The duration of the parallel
task ($T_g$) is measured when the worker ($S_n$) is complete:

\begin{equation}
T_g = \max\{S_1, S_2 ... S_m\}.
\label{eqn:max}
\end{equation}

These criteria broadly meet the preconditions for extreme value theory to be
applicable. And hence $T_g$ is approximated by $P(x)$ in Eqn
\ref{eqn:gevpdf}. The authors then present results for a carefully
constructed setup that attempts to minimize external perturbations and
focuses on solely on the period when the workload is running under IID
conditions.

This paper builds on work by studying a parallel workload in a production
HPC environment. The chosen workload is more typical to an actual workload
and the duration studied includes some serialized setup time. The
environment is a multi-user environment and no precautions have been taken
to limit any other simultaneous activity on the system. This paper argues
that the extreme distribution should follow a Fisher-Tippet Type I
(``Gumbel'') distribution. A framework for data analysis of extreme
statistics \cite{chapman01} is applied to distinguish between the different
Fisher-Tippett asymptotes.

\section{Experiment}

In changing from the controlled environment constructed in \cite{henwood16}
to a multi-user production HPC environment in this study, two notable
differences must be understood. Namely: traffic from external activates and
network topology.

\subsection{Traffic from external activities}

$T_g$ in Eqn \ref{eqn:max} is defined in \cite{henwood16} as the quiescent
performance of the file system under study. Let us simply assume that the
observed duration of a task on a system with congestion (from other jobs
accessing the shared resource) is the quiescent duration $T_q$ with
a delay of $T_s$ due to external congestion. i.e.

\begin{equation} T_g = T_s + T_q.
\label{basicvar}
\end{equation}
\begin{itemize}
\item $T_g$ is the observed write time of a fixed size transfer.
\item $T_s$ is the quiescent write time of a fixed size transfer.
\item $T_q$ is the time required to resolve resource contention on a multi-user system. 
\end{itemize}

By formalizing the affect of external congestion we now make the following
claim: provided the congestion factor ($T_s$) is constant throughout
the period of experimental observation, it can be ignored for the purposes of identifying the distribution present.

\subsection{Network topology}

From the perspective of the task in question, the worker nodes (storage
nodes in this case) are all assumed to be the same distance away. Depending
on the configuration of the HPC custer, this may, or may not be the case. In
this study, we chosen a HPC cluster (Ranger) is chosen where this
requirement typically holds.

The experiment has two parts. The first part is designed to measure the
distribution of a typical write. The second part of the experiment applies
the framework described by Chapman, Rowlands, and Watkins \cite{chapman01}
to verify the presence of a member of the GEV family with $\xi = 0$ in
Eqn. \ref{eqn:gevpdf}. With $\xi=0$, the GEV distribution would
specialize to the Gumbel distribution \cite{gumbel1958,coles2001}.

\subsection{Environment}

Experimental data were collected during the summer of 2012 on the Ranger HPC
environment \cite{minyard08experiences} operated by the Texas Advanced
Computer Center at The University of Texas at Austin. Ranger is a Sun
Constellation Linux cluster with system components connected via a full-CLOS
InfiniBand interconnect. Eighty-two compute racks house the quad-socket
compute infrastructure. A high-speed parallel Lustre file systems is
provided running across 72 I/O servers. 

The experimental runs were executed as normal user jobs during a normal
production operation. The parallel file system is a Lustre file system (v
1.8.5) on Red Hat$^*$ Enterprise Linux$^*$ 5, the client nodes are Red
Hat$^*$ Enterprise Linux$^*$ 5.

\subsection{Workload definition}

For the purposes of both identifying the underlying variability and testing
for particular member of the GEV family a typical write is defined as:

\begin{itemize}
\item Total file size is 16GB (half the available client memory.)
\item 16 storage nodes are used, a single 1GB stripe is written to each storage node.
\end{itemize}

A 1GB stripe size is chosen to ensure the effect of a serialized metadata
request (a restriction imposed by the Lustre 1.8 file system) is small and
the write time from the perspective of the client is dominated by variability
in the storage nodes. 16 nodes are chosen as a level of parallelism
sufficient to exhibit GEV behavior.

The tool {\tt{dd}} (v 5.2.1) is used to complete the write. A block size of
16GB is specified. A single directory is used for all writes.
{\tt{/dev/zero}} is used as the data for the write and is chosen to ensure
the client node does not stall waiting for data to write. During this phase,
the write is repeated 100 times to ensure sufficient fidelity of the
resulting distribution.

\begin{verbatim}
$LUSTRE = '/scratch/writetest' # directory
                # on the parallel file system
$SSZ = 1024 # each stripe is 1GB stripes
$SCT = 16 # number of stripes to write is 16
$FILESIZE = $SSZ * $SCT
lfs setstripe -c $SCT -s ${SSZ}M $LUSTRE
for i in {0..100} do; \
  sleep 30
  time dd if=/dev/zero of=$LUSTRE/${i}.dat \
                bs=${FILESIZE}M count=1 \
                >> ~/results.txt;
done
\end{verbatim}

To ensure 16 storage nodes are written in parallel, a Lustre control code
({\tt{lfs setstripe}}) is used to define the striping on the destination
folder. The client node enters sleep for 30 seconds between writes to mimic a
computation load that may be present during a typical application.  Once the
write is complete, {\tt{time}} diagnostics are written including the absolute
elapsed real time, which is the measure of interest $T_g$. Diagnostics are
written from the client to a separate file system. Note: the {\tt{dd}} tool on the Ranger environment is unable to write a block size of larger than 2GB. 
If a blocksize is specified larger than this limit, 2GB is written and the task ends.

\subsection{Testing for a specific member of the GEV family}

Chapman, Rowlands, and Watkins \cite{chapman01} provide a method for
identifying the presence of values distributed with the Gumbel distribution
(where $\xi=0$ in Eqn. \ref{eqn:gevpdf}) by exploiting the fact the value of
the third central moment (skewness) of the Gumbel distribution is constant
($\simeq 1.14$). The tools and setup are identical to the previous experimental
phase.  Instead of fixing on 16 storage nodes, the method requires a sweep of
increasing storage node counts.

The method is as follows: One hundred 1GB writes are completed against a single
storage node and the skewness calculated from the distribution of the elapsed
time measurements. The experiment is repeated, this time writing one GB each to
two storage nodes. Storage nodes and GBs are added until a limit is reached and
for each addition 100 writes are performed and the skewness is calculated for
the 100 experimental runs. In practice, {\tt{\$SCT}} is incremented in the code
example above.

As the number of storage nodes increases, the measured skewness of the
data should converge on the Gumbel skewness value. Chapman, Rowlands,
and Watkins \cite{chapman01} point out that the speed of the convergence
is dependent on the underlying distribution from which the maximum (or
minimum) is selected.

\section{Results}

\begin{figure}
\centering
\def\svgwidth{\columnwidth} 
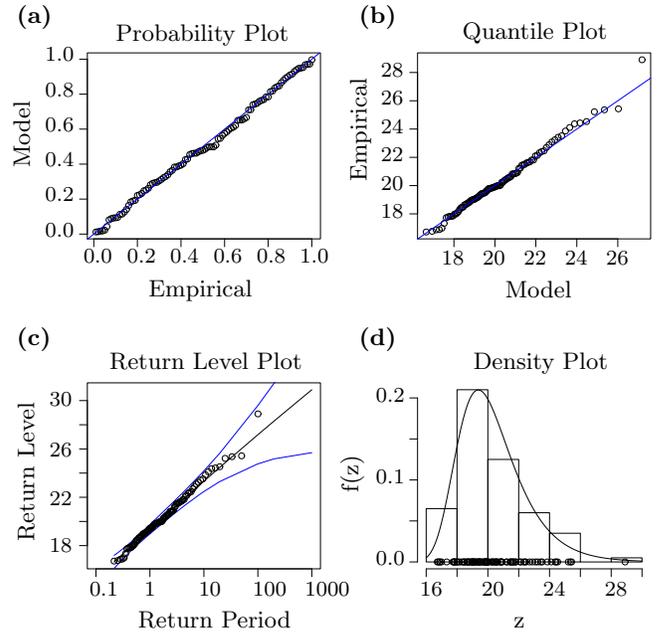
\caption{A model using the GEV distribution of a typical write against 16
  storage nodes repeated 100 times. 
A GEV distribution (Eqn. \ref{eqn:gevpdf}) is fitted to the data with a
  location $\mu = 19.36$ ($\pm 0.20$), scale $ \sigma = 1.75$ ($\pm 0.14$),
  and shape $\xi = -0.01$ ($\pm 0.07$).  {\bf{(a)}} shows the cumulative
  value of the observation against the model value. {\bf{(b)}} is the
  observed quantity plotted against the modeled quantity. {\bf{(c)}} shows
  the 95\% confidence interval of the value of $\xi$.  {\bf{(d)}} presents
  the observation histogram with the probability density plotted over the
  data.}
\label{fig:1nresults}

%
%
\end{figure}

Fig.\ref{fig:1nresults} shows the distribution of 100 typical writes
designed to deliver a load characteristic of large file count per directory
large I/O's checkpoint. The plot {\bf{(a)}} focuses attention to the state
of the fit around the distribution's head. The quantile plot {\bf{(b)}}
emphasises the tail of the fit. Points lying on the line indicate perfect
agreement between observation and model. Both the probability plot
{\bf{(a)}} and the quantile plot {\bf{(b)}} show good agreement with the
model.  The return level plot {\bf{(c)}} also indicates good agreement with
the model as the points appear within the 95\% confidence level.  {\bf{(d)}}
provides further evidence showing the fitted GEV density function is in good
agreement with the data histogram.  All figures was generated using the R
language \cite{rmanual} using the ismev library \cite{ismev}.

\begin{figure}
\centering
\def\svgwidth{\columnwidth} 
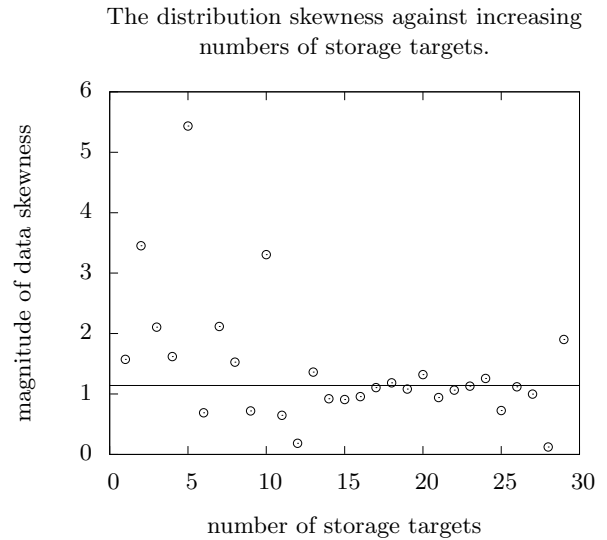

\caption{The skewness of the individual distributions obtained by increasing
  the number of storage nodes in the system. For each run with a fixed number
  of storage nodes, 100 typical writes were completed. The skewness of the
  Gumbel distribution is shown as a horizontal line at 1.14.}

\label{fig:addostsresults}
\end{figure}

Fig.\ref{fig:addostsresults} skewness and mean of write times of the
individual distributions obtained as the number of storage nodes is
increased.  A large spread of distributions is present as storage nodes
increase from one to 12.  Between 12 and 27 the data skewness is near the
value for the Gumbel distribution (1.14).  Beyond 27 storage nodes, the
skewness of the data appears to spread away from the Gumbel skewness.

The maximum number of storage nodes included within this experiment was
limited to 29. Beyond this count, the client node exhausted local memory and
did not successfully complete the write.

\section{Discussion}

The results obtained from the Ranger system are encouraging. They were
obtained with a simple experiment, on a production system running many
simultaneous jobs. 

Fig.\ref{fig:1nresults} shows general agreement with the GEV model.  The
value of the shape parameter $\xi = -0.01 \pm 0.07$ is close to the shape
value for the Gumbel distribution ($\xi = 0$.) Observations within the tail
may appear to diverge slightly from the model, but remain well within the
margin of error (95\%).

If we consider Amhdal's Law:

\begin{equation}
  S_{latency}(s) = {{1}\over{(1-p) + {{p}\over{s}}}}
\label{eqn:amdahl}
\end{equation}

This equation give the theoretical speedup in latency of a parallel task
given the proportion of the task that executes in parallel. In the previous
study of GEV in parallel environments, the proportion of the application that
executed in parallel was made as close to 100\% ($p = 1$) as possible. With
the use of {\tt{dd}} to dispatch the parallel workload in study, there is a
non-zero period of the task spent performing the serial metadata operations
before the parallel work begins. This suggest that GEV varibility is also
present for workloads where $p \approx 1$.

Fig.\ref{fig:addostsresults} indicates a rapid convergence towards the
Gumbel distribution from one to 12 storage nodes. Between 12 and 27 storage
nodes the skewness of the distribution is close to the Gumbel skewness.
Beyond that value, the skewness is less like Gumbel. 

One possible explanation for the trend away from Gumbel skewness with higher
storage node counts is congestion from external traffic. As the storage node
count increases on a multi-user system, it is much less likely the
assumption to reduce $T_q \rightarrow 0$ in Eqn. \ref{basicvar} will hold.
i.e., while there are a large number of storage nodes available, as the
client writes to an increasing number, the probability another job will also
be requesting the same storage node resources increases. If a storage node
is under higher load it may service requests with a distribution that is not
identical to an unloaded storage node.

\begin{figure}
\centering
\def\svgwidth{\columnwidth} 
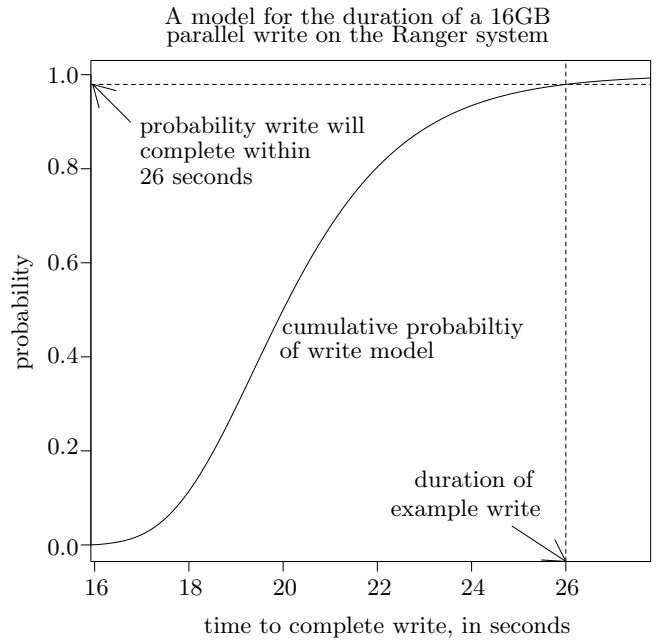

\caption{A model of a write on a parallel file system to 16 storage nodes.
  The model is a cumulative generalised extreme distribution function with
  values of location $\mu = 19.36$ ($\pm 0.20$), scale $ \sigma = 1.75$
  ($\pm 0.14$), and shape $\xi = -0.01$ ($\pm 0.07$). The probability a
  write will be completed in 26 seconds or less is illustrated graphically.
  The probable value is about 98\%.}

\label{fig:gevmodel}
\end{figure}

Fig.\ref{fig:gevmodel} presents the cumulative distribution of the GEV model.
The model uses the parameters extracted from the fit in Fig.
\ref{fig:1nresults}. A graphical technique for classifying the likelihood that
a particular write will take a given duration is shown with dotted lines. In
this case, the example write is observed to take 26 seconds. A vertical line is
drawn from the measure of 26 on the abscissa to intersect the cumulative
distribution function. From this point of intersection, horizontal line is
drawn to the ordinate intersecting at a probability of about 98\%.

The result in the example presented in Fig.\ref{fig:gevmodel} is that the
probability a write will take 26 seconds or less is about 98\%. In
contrast, writes that take more than 26 seconds have a probability of about
2\% - or occur approximately once for every 50 writes. Given this, if a
client observes more than one write in 50 taking longer than 26 seconds an
alert could be issued.

\section{Conclusions}

Understanding and predicting the behavior of high performance computers is a
challenging task. Even after they are constructed and operating the
measurement, benchmarking, and monitoring of large computers is still an
art, practiced by a small number of experts. This paper presents extreme
value theory as a new tool to enhance diagnostics. Taking the required
measurements to reveal the I/O distribution for a given machine is shown
here to be simple and quick to complete. Performing such measurements on a
quiet system (where $T_q = 0$) enables the baseline variation of the
environment to be captured.  The model presented in this paper can be
calibrated for a given environment and periodic observations can reassure a
systems operator that the machine is performing optimally. Furthermore,
depending on the operations policy of the environment an accurate model for
variability may be valuable in pricing service level agreements.

It is interesting to consider a practical outcome of the GEV result for
increasing parallel-ism in data transfers: the case where the throughput of the
client interface is the narrowest (or equally narrow) bottle neck in the
system. In this case, a parallel transfer is constricted at the point of the
client. Adding parallelism will have the effect of reducing performance, and
increasing variability. Hence, one might conclude that, if possible, a single
storage target should be specified for all data transfers.

As high performance computing continues to develop, new libraries become
available to simplify interfacing with data objects. For example, the
{\tt{t3pio}} library provides automatic configuration for MPI applications
that use HDF5. Such a library could introduce auto-tuning behavior to avoid
congestion from competing file system tasks if the assumption that a write
time will have a GEV distribution. A single client node could be calibrated
for ideal parallel behavior and measure deviations from this behavior as
values that are unlikely according to the GEV model. A more sophisticated
client could choose to modify file layout or cross-reference with other
anomalous results to identify congestion areas within the system.

Predicting the behavior of a large machine while it is still in design is
valuable if accurate. The GEV model for I/O enables a designer to sweep away
the complexity of modeling the file system as a series of components.  By
simulating the parallel file system as a simple GEV (or Gumbel) source it may
be possible to arrive at an accurate file system performance metric. It is
conceivable that by starting with the specified performance of a single disk
and successively applying the GEV model one can now arrive at the probable
behavior of the whole storage system.

\section{Acknowledgements}

RH is grateful to John Hammond, Doug James and Andreas Dilger for sharing
their expertise on the subtleties of a Lustre file system on Ranger. Linda
Berbernes and Victor Eijkhout provided valuable input to enhance the text and
content.  SCC acknowledges a Fulbright Lloyd's of London Scholarship and 
support from the AFOSR on grant FA9550-17-1-0054.

\bibliographystyle{abbrv}
\bibliography{refer}

\end{document}